\documentclass[graybox]{svmult}

\usepackage{type1cm}        
\usepackage{makeidx}         
\usepackage{graphicx}        
\usepackage{multicol}        
\usepackage[bottom]{footmisc}

\usepackage{newtxtext}       %
\usepackage[varvw]{newtxmath}       

\usepackage{adjustbox}
\usepackage{graphicx}
\usepackage{multirow}
\usepackage{float}
\usepackage{longtable}
\usepackage{tabularx}
\usepackage{hhline}
\usepackage{booktabs} 

\usepackage{url}
\urldef{\mailsa}\path|{alfred.hofmann, ursula.barth, ingrid.haas, frank.holzwarth,|
\urldef{\mailsb}\path|anna.kramer, leonie.kunz, christine.reiss, nicole.sator,|
\urldef{\mailsc}\path|erika.siebert-cole, peter.strasser, lncs}@springer.com|    

\makeindex             

\begin{document}

\title*{Autonomic Cloud Computing: Research Perspective}
\author{Sukhpal Singh Gill\orcidID{0000-0002-3913-0369}}
\institute{Sukhpal Singh Gill \at School of Electronic Engineering and Computer Science, \\ Queen Mary University of London, UK, E14NS. \\ \email{s.s.gill@qmul.ac.uk}}
%
%
\maketitle

\abstract{As the cloud infrastructure grows, it becomes more challenging to manage resources in such a massive, diverse, and distributed setting, despite the fact that cloud computing provides computational capabilities on-demand. Due to resource variability and unpredictability, resource allocation issues arise in a cloud setting. A Quality of Service (QoS) based autonomic resource management strategy automates resource management, delivering trustworthy, dependable, and cost-effective cloud services that efficiently execute workloads. Autonomic cloud computing aims to understand how computing systems may autonomously accomplish user-specified "control" objectives without the need for an administrator and without violating the Service Level Agreement (SLA) in a dynamic cloud computing environments. This chapter presents a research perspective and analysis on autonomic  resource allocation in cloud computing based on the last decade of conducted research with a focus on QoS and SLA-aware autonomic  resource management. This study delves into the current state of autonomic resource management in the cloud and introduces a conceptual model for Artificial Intelligence (AI)-driven autonomic cloud computing. This model aims to optimise server load distribution and energy consumption, thus enhancing cost savings and environmental impact. Finally, it highlights key next-generation research directions.}

\keywords{Cloud Computing, Autonomic Cloud Computing, Artificial Intelligence, Autonomic Computing, Service Level Agreement, Quality of Service}

\section{Introduction}
Cloud computing is an evolving utility computing mechanism in which cloud consumers can detect, choose, and utilize the resources (infrastructure, software and platform) and provide service to users based on a pay-per-use model as computing utilities \cite{A1}. The current computing mechanism is effective, particularly for medium and small cloud based companies, in that it permits easy and reliable access to cloud services like infrastructure, software, and platforms. Current cloud computing bears a strong resemblance to the established models of cluster computing and grid computing \cite{A2}. The important key technical features of cloud computing include autonomic service, rapid elasticity, end-to-end virtualization support, on-demand resource pooling, and transparency in cloud billing. Further, non-technical features of cloud computing include environment friendliness, little overhead maintenance, lower upfront costs, faster deployment times, Service Level Agreement (SLA) and pay-as-you-go- models \cite{A3}. In a distributed computing environment, the unpredictability of services is a fact, so the same possible in the cloud as well. The success of next-generation cloud computing infrastructures will depend on how capably these infrastructures will discover and dynamically tolerate computing platforms, that meet the randomly varying resource and service requirements of cloud customer applications  \cite{A4}.

\section{Service Level Agreements (SLAs)}
SLA designates what you require from your consumers/service customers in order to provide the service specified. It needs assurance and support from both parties to provision and follow the contract in order for the SLA to work efficiently  \cite{A5}. In the SLA, both parties (the cloud provider and the cloud consumer) should have specified the possible deviations to achieve appropriate quality attributes. In other words, it is vital for users to acquire assurances from suppliers on service provisions. Typically, providers and customers negotiate SLAs to deliver these assurances. The very first problem is the description of SLA terms in such a way that has a suitable level of granularity, namely the compromises between accuracy and complexity, so that they can ensure most of the user hopes and are comparatively simple to be prejudiced, certified, calculated, and imposed by the resource provisioning mechanism on the cloud  \cite{A3}. In addition, different cloud service models (IaaS (infrastructure as a service), PaaS (platform as a service), and SaaS (software as a service)) will need to express different SLA meta disclaimers  \cite{A6}. This also increases the number of implementation issues for cloud providers. Moreover, innovative SLA mechanisms require continuous integration of consumer response and customization features into the SLA assessment framework. As cloud service models develop and become omnipresent, it increases the probability of clarifying the way the services are provisioned and managed. As a result, it allows providers to address the various needs of their customers  \cite{A7}. In this perspective, SLAs appear as a significant characteristic that subsequently serves as the basis for the predictable quality level of the services made available to customers by the providers. Nonetheless, the collection of the recommended SLAs by providers (with marginal overlaps), has led to manifold different definitions of cloud SLAs  \cite{A8}. Moreover, confusion exists as to what the difference is between SLAs and agreements, what the marginal quality is, what terms are involved in each one of these documents, and if and how they are associated.

\begin{figure*}[t]
	\centering
	\includegraphics[scale=0.8]{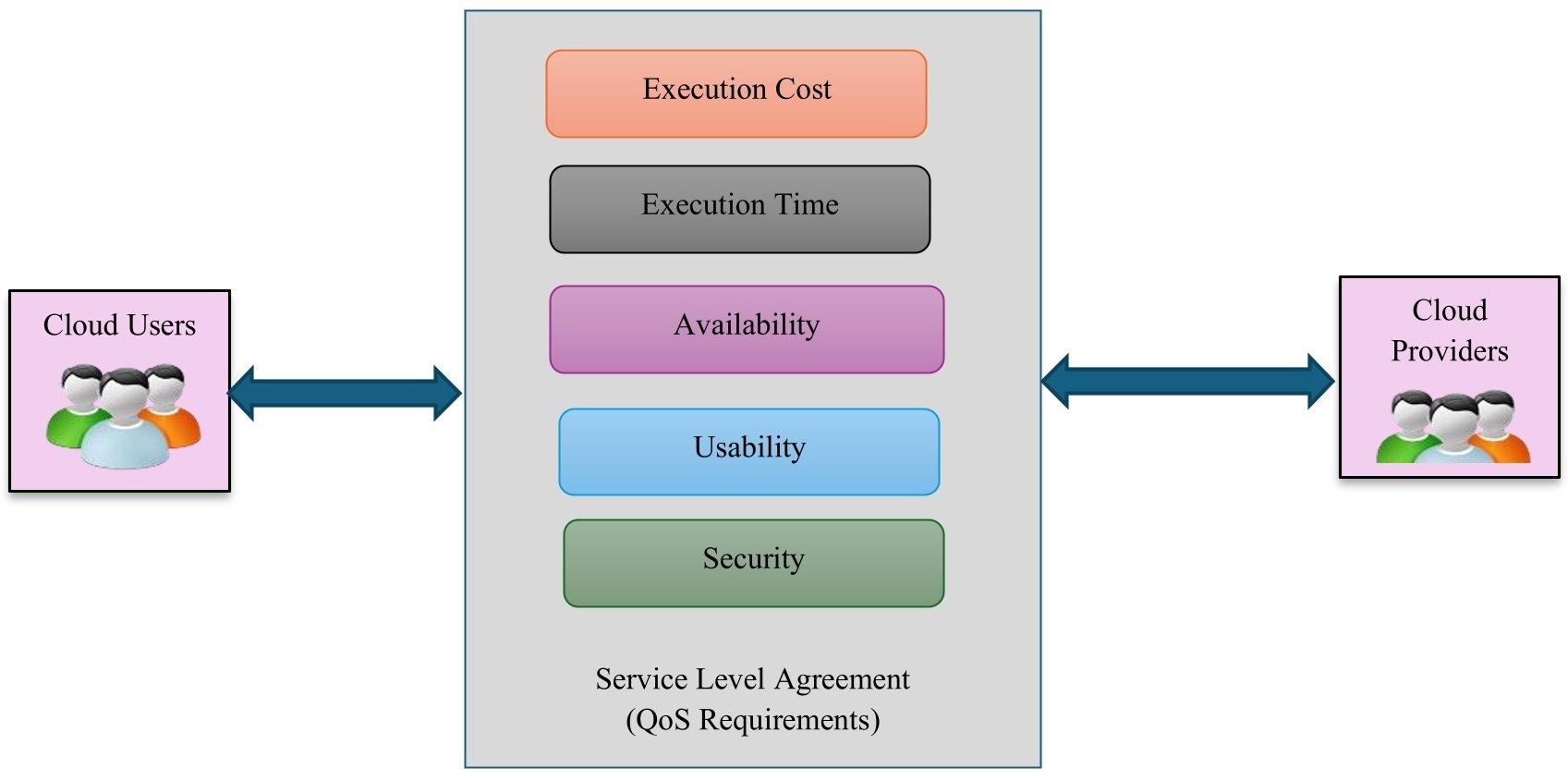}
	\caption{SLA Negotiation between Cloud Users and Cloud Providers \cite{A23} \cite{A38}}
	\label{fig1}
\end{figure*} 

\section{SLA and QoS in Cloud}
Logically, based on QoS requirements such as scalability, high availability, trust, and security, these applications will be characterized, identified in the so-called SLAs   \cite{A4}. Figure \ref{fig1} shows the process of SLA negotiation between cloud users and cloud providers. The current cloud technology is not completely personalized to honor probable SLAs, though both industrial and academic research groups are presenting increasing interest in problems of QoS assurance within the context of cloud computing  \cite{A9}. In general, an SLA necessitates a precise assessment of the characteristics of the required resources. Application services introduced in clouds (e.g., web applications, web services) are frequently characterized by great load inconsistency; therefore, the amount of resources required to honor their SLAs may vary, particularly over time. An important challenge for cloud providers is to automate the management of virtual servers while keeping into account both the high-level QoS requirements of hosted applications and resource supervision expenses \cite{A10}. Cloud market mechanisms are consistently static and cannot react to dynamic variations in consumer desires. To respond to these issues, there is a requirement for an adaptive methodology for autonomically springing SLA patterns based on consumer requirements. The present research in cloud SLA limits the capability of matching conformation metrics to acceptable. These metrics comprise statistical measures such as the standard deviation that want to be computed from the expected and actual outcomes of services delivered to customers \cite{A11}. Semantic Web technologies can be used to improve the descriptions and, therefore, increase the quality of these matches. Although cloud consumers do not have full supervisory control over the fundamental computing resources, they do require ensuring attributes such as quality, accessibility, trustworthiness, and performance of these resources when users have transferred their fundamental business functions onto their honoured cloud \cite{A12}.

\begin{figure*}[t]
	\centering
	\includegraphics[scale=0.9]{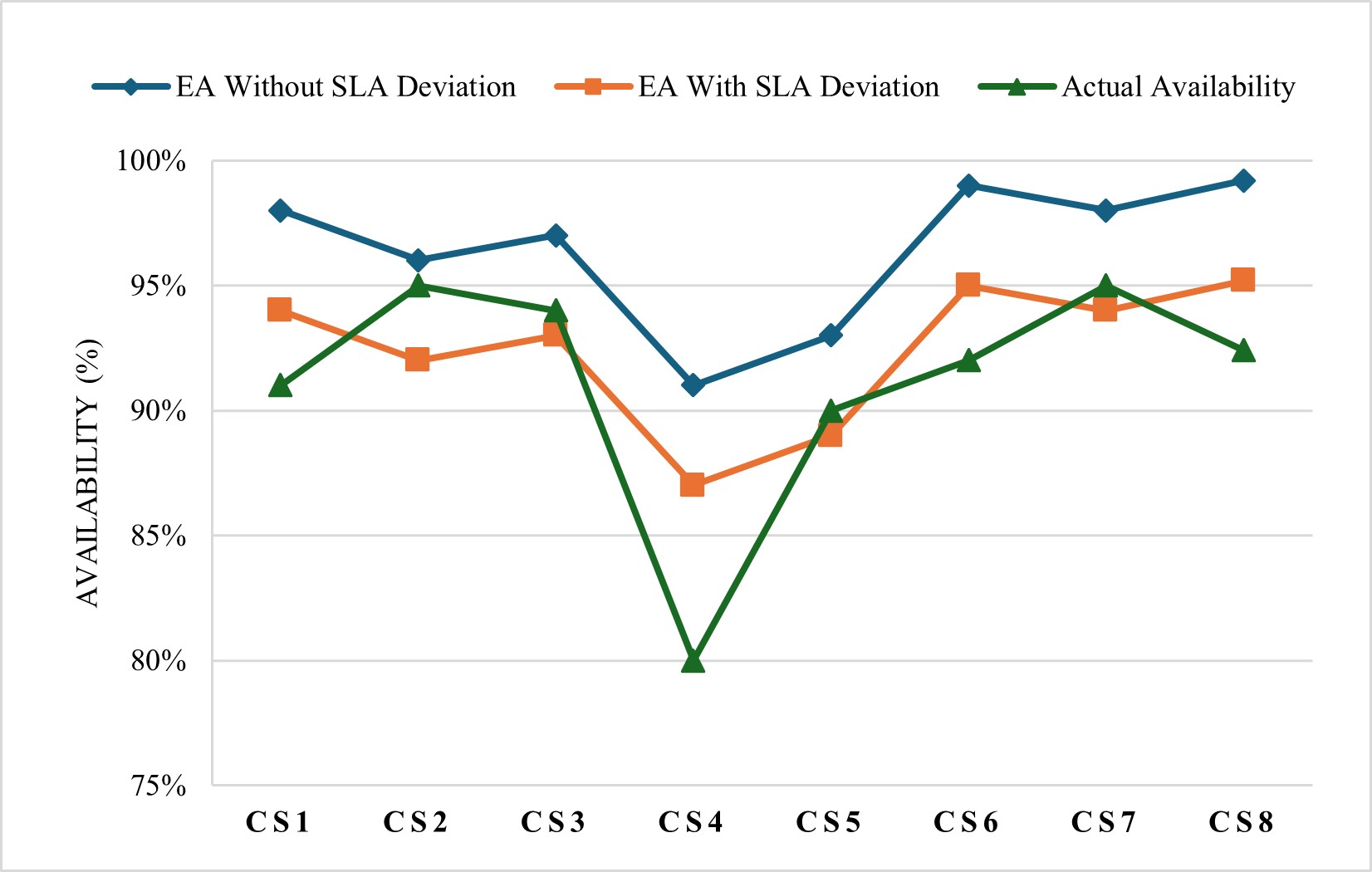}
	\caption{Comparison of SLA deviations \cite{A20} \cite{A23}}
	\label{fig2}
\end{figure*} 

\section{Analysis of SLA Deviation}

\begin{figure*}[t]
	\centering
	\includegraphics[scale=.9]{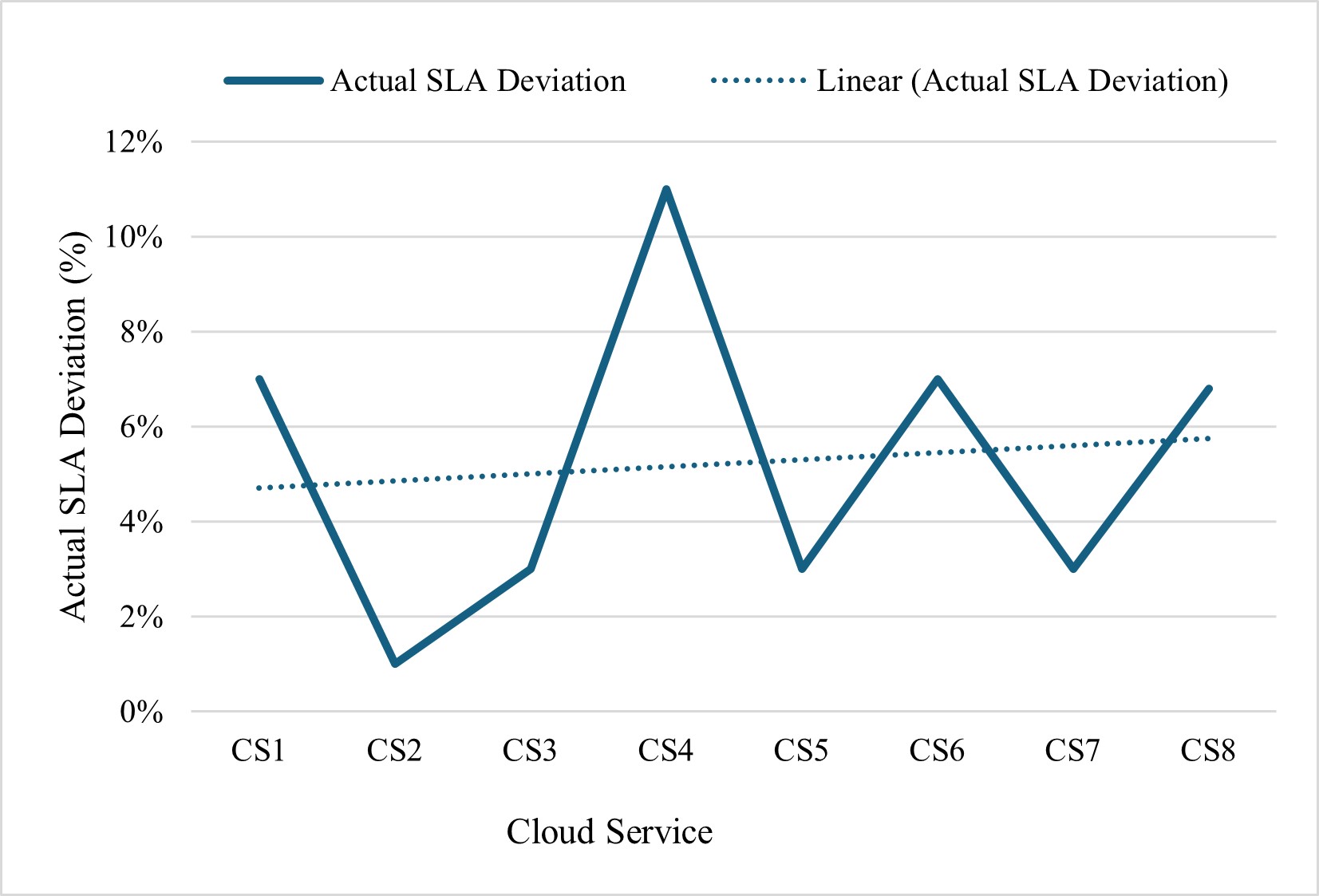}
	\caption{Actual SLA Deviation  \cite{A20} \cite{A23}  \cite{A23}}
	\label{fig3}
\end{figure*}

Assume that there are 8 different cloud services being provided by the cloud provider with the same SLA deviation (4\%) for each service by taking availability as a quality attribute, as shown in Table \ref{table1}. Table \ref{table1} describes the Expected Availability (EA) without SLA deviation, Expected Availability (EA) with SLA deviation, Actual Availability (AA), Actual SLA deviation and whether the SLA violates for a particular service or not. The comparison of Actual deviation (vary) and Expected deviation (4\%) of 8 cloud services is shown in Figure \ref{fig2}. Figure \ref{fig3} shows the comparison of SLA deviations for eight cloud services. 

\begin{table*}[ht]
\centering
\small
\caption{Cloud SLA Deviations  \cite{A20} \cite{A23}}
\begin{adjustbox}{angle=0}
\resizebox{0.85\textwidth}{!}{
\begin{tabular}
{p{0.5in}|p{1in}|p{1in}|p{.7in}|p{.7in}|p{0.5in}}
   
\hline
\textbf{Cloud Service} & \textbf{Expected Availability  Without SLA Deviation} & \textbf{Expected Availability  With SLA Deviation} & \textbf{Actual Availability} & \textbf{Actual SLA Deviation} & \textbf{SLA Violates} \\ \hline
CS1                    & 98\%                              & 94\%                           & 91\%        & 7\%                           & Yes                   \\ \hline
CS2                    & 96\%                              & 92\%                           & 95\%        & 1\%                           & No                    \\ \hline
CS3                    & 97\%                              & 93\%                           & 94\%        & 3\%                           & No                    \\ \hline
CS4                    & 91\%                              & 87\%                           & 80\%        & 11\%                          & Yes                   \\ \hline
CS5                    & 93\%                              & 89\%                           & 90\%        & 3\%                           & No                    \\ \hline
CS6                    & 99\%                              & 95\%                           & 92\%        & 7\%                           & Yes                   \\ \hline
CS7                    & 98\%                              & 94\%                           & 95\%        & 3\%                           & No                    \\ \hline
CS8                    & 99.2\%                            & 95.2\%                         & 92.4\%      & 6.8\%                         & Yes                   \\ \hline
\end{tabular}
}
\end{adjustbox}
\label{table1}
\end{table*}

\section{Existing Challenges in Traditional Cloud Computing}
Usually, cloud providers do not consider compensation because resource providers do not actually enforce penalties for SLA violations. There should be a penalty delay cost or consumers’ compensation if the cloud provider misses the deadline  \cite{A13}. Furthermore, when the cloud provider violates the terms, it transfers the risk to IaaS providers. Penalty delay cost is how much the service provider has to give concessions to users for SLA violations. It is dependent on the penalty rate and the penalty delay time period   \cite{A14}. Researchers have tried to reduce the effect of inaccuracy using two approaches: first, considering the penalty compensation clause in SLAs with IaaS providers and imposing SLA violations; and second, adding some slack time during scheduling to avoid risk  \cite{A15}. However, this increases the cost and execution time significantly  \cite{A11}  \cite{A12}. Autonomic cloud computing  can be utilized to solve this problem, understanding how computing systems can autonomously accomplish user-specified "control" objectives without the need for an administrator, without compromising QoS, and without violating the SLA in dynamic cloud computing environments.

\section{Potential Solution: Autonomic Cloud Computing}

The Autonomic Computing Initiatives (ACI) at IBM became one of the earliest sector-wide programmes to focus on designing computers to work well with no or minimal intervention by humans  \cite{A16}. IBM's Tivoli Frameworks group originally tested autonomic computing to optimise DB2 efficiency. The endeavour had an enormous impact on studies of how the nervous systems of humans work together. For example, how the autonomic nervous system responds to stimulation even when the user isn't consciously aware of it and how an autonomic computing environment uses advanced computational intelligence without the user even realising it's happening  \cite{A17}. Additionally, humans' nervous systems can do many simple things at the same time, like change the internal environment, change the rate of breathing, and release chemicals through pores in response to stimuli. All of this happens while following rules and "limits" and based on signals we receive or sense from our surroundings or from within ourselves  \cite{A16}. An autonomous technology ecosystem operates on data collected, sensed, or taught, akin to how the human body processes information, without requiring immediate human oversight over system management operations.

\subsection{Background}
Biological networks are a source of inspiration for QoS-based autonomous systems because of their remarkable ability to deal with unpredictability, variability, change, errors, and so on. Executing a programme within a deadline while satisfying user-described QoS standards with minimal complexity is the purpose of autonomic systems  \cite{A17}. The human's autonomic nervous system (ANS) is a primary model for autonomic  computing systems. When faced with a volatile environment, the ANS can dynamically handle and manage any circumstance that may arise  \cite{A23}. Similar to how the ANS regulates bodily activities like respiration and digestion, autonomic  cloud computing systems (ACCSs) automate the operation of cloud-based software and systems. In the same way that the ANS checks and monitors, ACCSs react situationally by doing things such as \textit{self-optimizing, self-configuring, self-protecting,} and \textit{self-healing} \cite{A18}.

\subsection{Self-* Properties}
\begin{figure*}[t]
	\centering
	\includegraphics[scale=.9]{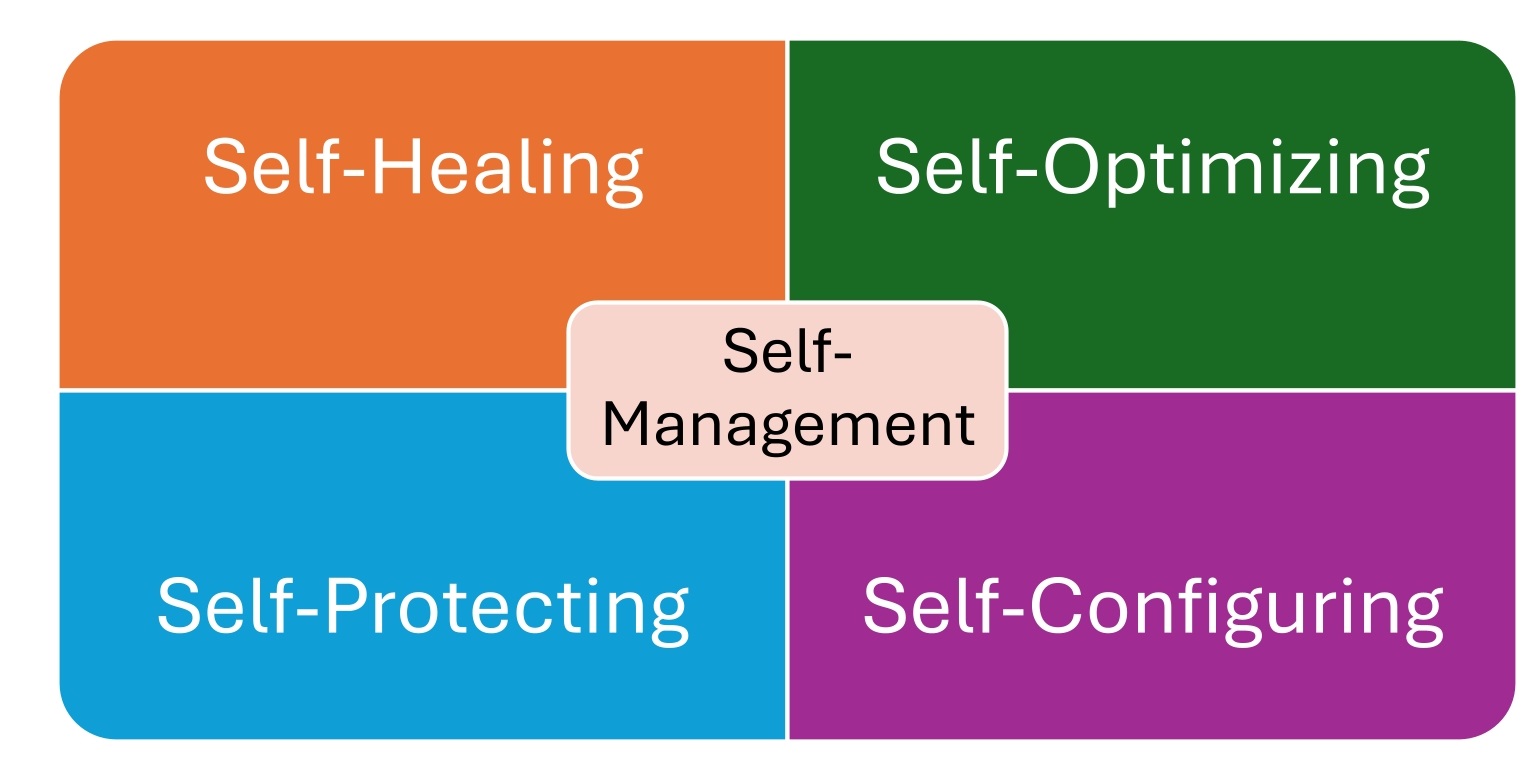}
	\caption{Self-* Properties of Autonomic Computing System \cite{A16} \cite{A38}  \cite{A23}}
	\label{fig4}
\end{figure*}

Autonomic computing, also called self-adaptation mechanisms, is an area of study that focuses on how computers may independently accomplish preferred actions. "Self-*" systems are a popular way to describe these kinds of systems; the "*" indicates the sort of actions, such as Self-"configuration, optimization, protection, healing"  \cite{A19}. Figure \ref{fig4} shows the four properties of autonomic computing system. 

A \textit{self-configuring} autonomic system is one that can adjust to new conditions without external interference  \cite{A22}. Based on error signals or warnings issued by a monitoring device, the framework repairs faulty or outdated elements autonomously  \cite{A20}. Autonomic systems that can optimise themselves by decreasing resource overload and under-utilisation and effectively performing computing tasks are called \textit{self-optimizing}  \cite{A18}. An autonomic system's ability to \textit{self-protect} means it is capable of warding off intruders and hackers  \cite{A19}. The system's autonomic controller must additionally be able to recognise and avoid damaging attacks. An autonomic  system can self-diagnose, self-evaluate, and self-repair when it makes mistakes; no human interaction is required for this process, referred to as \textit{self-healing}  \cite{A22}. This self-* attribute improves performance by reducing the impact of errors on execution, a feature known as fault tolerance. In an real world, self-managing and self-healing systems would never require human intervention for setup or maintenance. When taken as a whole, the above-mentioned characteristics must fall under the supervision of self-managed systems  \cite{A21}.

\subsection{Architecture}
\begin{figure*}[t]
	\centering
	\includegraphics[scale=.7]{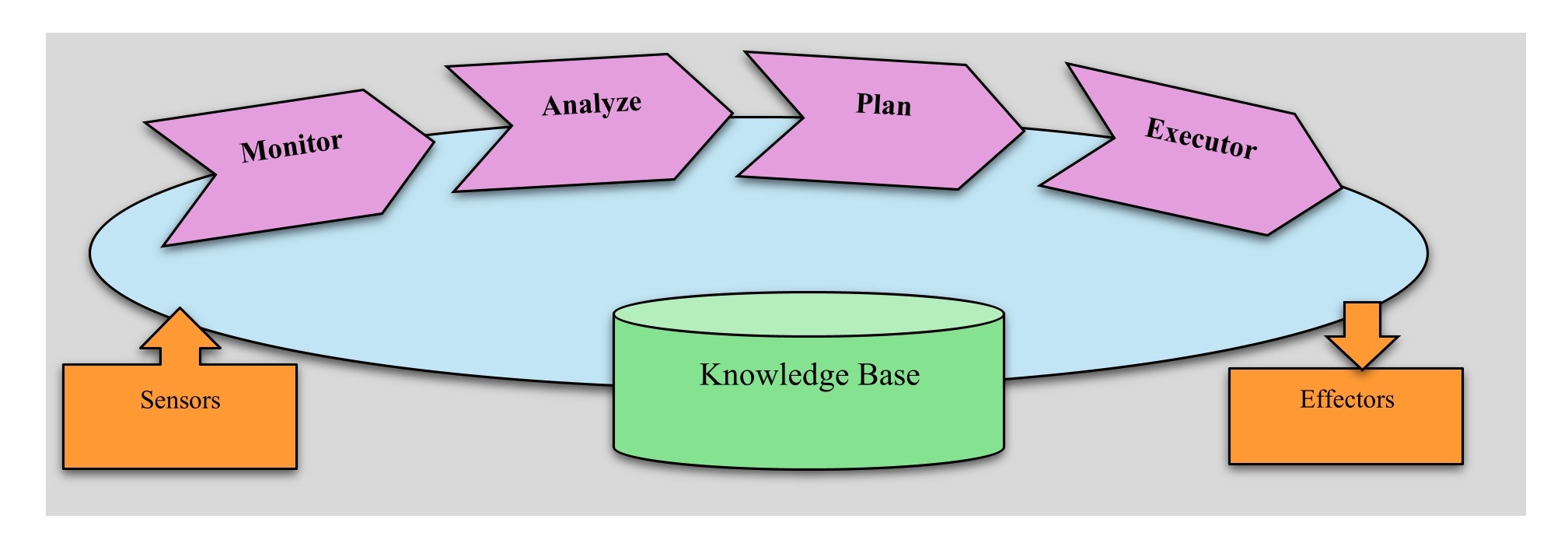}
	\caption{General Architecture of Autonomic Computing System \cite{A16}  \cite{A38}  \cite{A23}}
	\label{fig5}
\end{figure*}

Various real-world systems accomplish these goals with varied degrees of precision and effectiveness. The amount of oversight and involvement from humans can also differ. The Autonomic Manager (AM) is an intelligent unit that operates under IBM's Autonomic Computing architecture \cite{A16}. It communicates with its surroundings through administrative graphical user interfaces such as \textit{sensors} and \textit{effectors} and takes action according to data gathered from sensors and policies stored in a low-level database. Using broad-brush notifications and directives, an administrator sets up the AM. IBM implements the autonomic process, as illustrated in Figure \ref{fig5}. Autonomic systems are frequently organised as monitor, analyse, plan, and execute (MAPE) phases.

First, \textit{monitors} interact with external hardware to obtain sensor data, which they then send to the next component for further analysis in order to regularly review QoS measures  \cite{A23}. The monitoring module's data is evaluated in the \textit{Analyse} and \textit{Plan} modules, which then formulate suitable responses to system alarms. This autonomic system responds appropriately to the alerts issued based on the data analysis results. The execution plan has to be put into effect by the \textit{executor}, and its key purpose is to preserve the QoS of an execution programme after a comprehensive assessment involving verifying and validating it to offer assurance that the modification will truly operate  \cite{A24}. An executor monitors the addition of new information to the database and responds to the results.

\begin{figure*}[t]
	\centering
	\includegraphics[scale=.9]{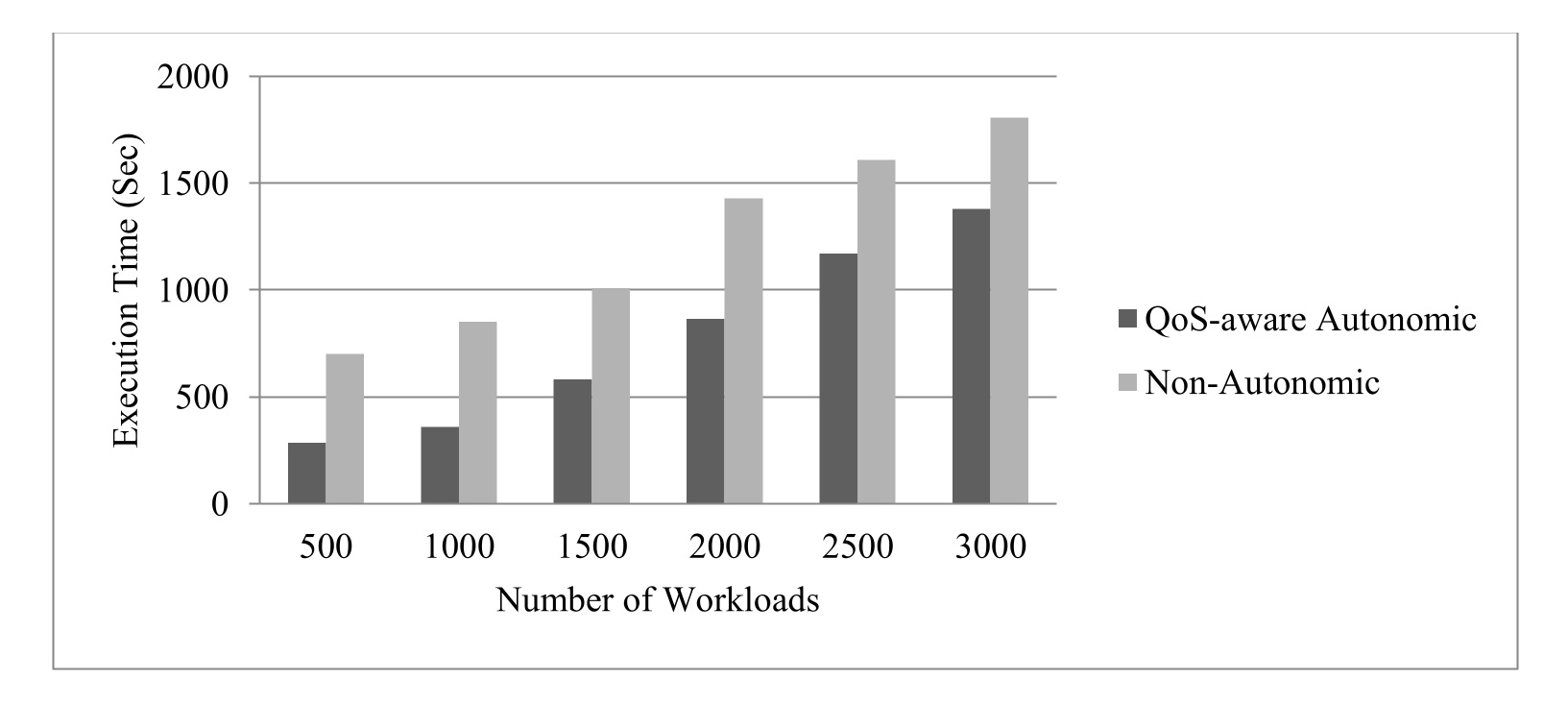}
	\caption{Comparison of Autonomic and Non-Autonomic Computing Systems based on QoS (execution time) \cite{A20} \cite{A24} \cite{A40}}
	\label{fig6}
\end{figure*} 

\begin{figure*}[t]
	\centering
	\includegraphics[scale=.9]{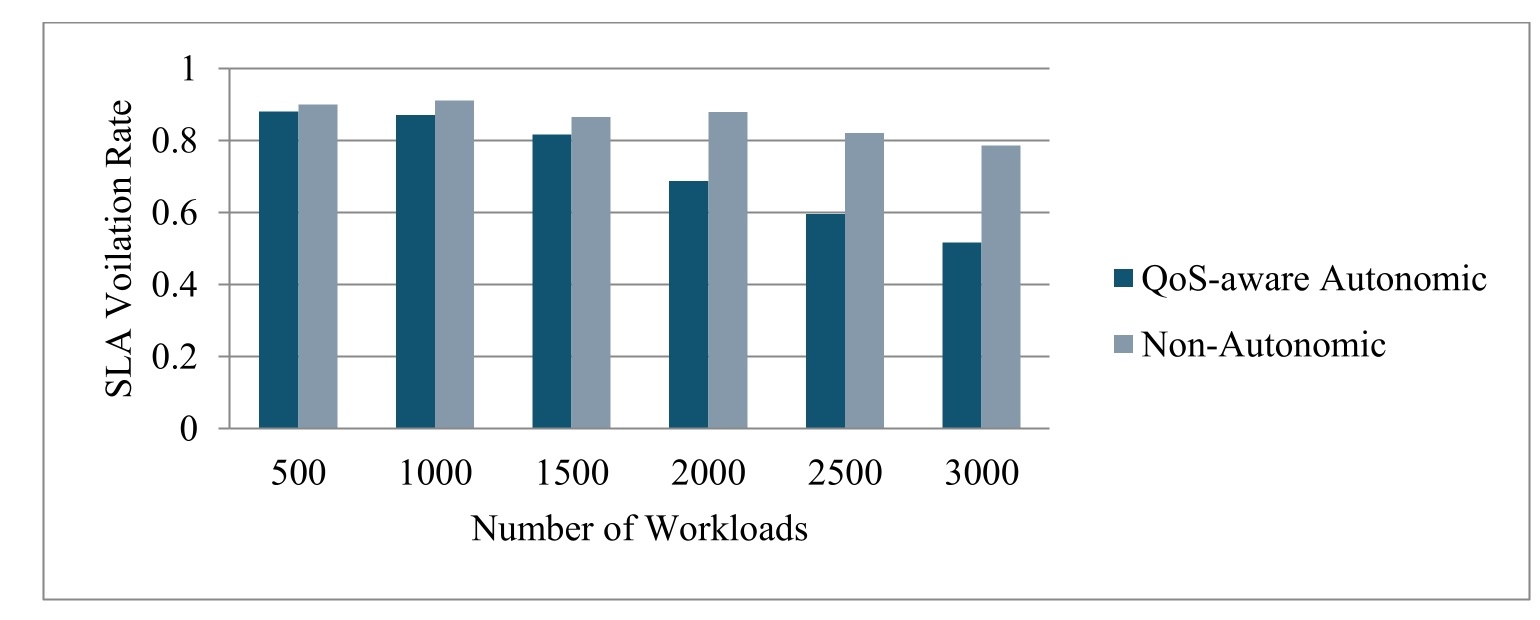}
	\caption{Comparison of Autonomic and Non-Autonomic Computing Systems based on SLA violation rate \cite{A20} \cite{A24} \cite{A40}}
	\label{fig7}
\end{figure*} 

\subsection{Validation}
Based on QoS characteristics, smart autonomic systems take cues from biological systems that have proven proficient at dealing with challenges the include dynamism, diversity, and the unpredictability of cloud resources. A smart autonomous framework aims to deploy applications on time while meeting the user-described QoS requirements with minimal complexity. The MAPE-K loop is the basis of an autonomic  computing system, which takes into account all four phases of an autonomic system with a control loop: monitoring, analysing, planning, and execution \cite{A16}. The simulation-based experiments have been conducted using the CloudSim toolkit  \cite{A25} to verify the use of autonomic computing for cloud resource management, as demonstrated in notable previous research  \cite{A19}  \cite{A20}  \cite{A21}  \cite{A22}  \cite{A23}  \cite{A24}. Autonomic systems outperformed non-autonomic computing systems experimentally in terms of QoS (execution time) and SLA violation rate, as shown in Figure \ref{fig6} and Figure \ref{fig7}, respectively, because the service can self-manage according to its environment's demands. Through the implementation of autonomic computing, application utilisation and efficiency were maximised, while performance and other quality of service assurances were ensured  \cite{A20}. These results prove without a reasonable doubt that autonomic cloud computing maintains system stability in the face of uncertainty and allows for rapid adaptation to novel environmental situations, such as software or hardware faults  \cite{A21}.

\section{AI and Computing: Intertwined in Autonomic Cloud Computing }
Using data obtained on system functions, Artificial Intelligence (AI) and Machine Learning (ML) could potentially be employed to assist with and generate autonomic behaviours. For instance, ML methods may be employed to identify trends within the workload, which can then be utilised to enhance the administration of available resources  \cite{A26}. The autonomic manager might also automatically call upon ML-based, continually changing system recognition approaches such as neural networks to do self-learning, which would help reduce modelling uncertainty. This allows for the generation of both black-box and grey-box models of the controlled systems throughout a concept movement, which can then be checked for sanity or, worse, for the detection of extremely important changes to the system's functioning. Autonomic systems are frequently organised as MAPE phases, and AI may be used in conjunction with management approaches to aid in the analysis and planning parts of these phases  \cite{A27}. Important advantages of autonomic self-management come from combining feedback control with data-driven model creation using ML, which provides important advantages for autonomic self-management.

Several computing frameworks, including cloud, fog, edge, serverless computing, and quantum technology, have included autonomic  computing using AI and ML methods  \cite{A28}. Whenever a system has many possible settings, autonomic computing approaches become very important. Optimising search across a larger number of possible alternatives is achievable when there is a larger potential range of parameters in which preferences can change. The best use case for autonomic  computing approaches is in the background, as a programmable gateway that applications can directly call   \cite{A29}.

Due to the inherent autonomy of most Peer-to-Peer (P2P) networks, there are several apps that can handle node failures, network setup and updates, and, to a lesser extent, optimise performance independently. More and more online services and data centre management tools are incorporating AI and ML-based self-management abilities \cite{A30}. This allows these platforms to swiftly adapt to changing demands. Schedulers and workflow administrators don't always have autonomic characteristics because they can't keep tabs on the system's health and offer real-time feedback, which makes them less autonomous   \cite{A31}. One way to increase these systems' capabilities is to incorporate "tuning" capabilities that utilise AI/ML approaches. Hadoop and MapReduce are two examples of self-managed computing systems that offer self-organization and self-healing features that make extensive resource utilisation possible.
\subsection{Transformative Effects of AI on Autonomic Cloud Computing}
As systems grow in size and connectivity, AI- and ML-based autonomous computing will become the norm, making the manual handling and customisation of these systems difficult and costly. It has been anticipated that AI and ML will primarily drive autonomic computing in the coming years, while humans will still have the ability to influence these systems' actions through carefully designed interfaces. Cyberphysical technologies and digital twins will make quality-assured and mission-critical adaptations imperative   \cite{A39}. This is why autonomous programmes will be accountable for physical resources, especially processing facility functions.

AI-powered autonomic computing significantly reduces overall maintenance costs. Consequently, spending on maintenance will go down. The number of individuals required to keep the systems running will also decrease. AI-powered autonomous IT systems provide benefits such as reduced time and money spent on installation and upkeep, as well as improved system reliability   \cite{A32}. The higher-order benefits state that businesses will have an easier time managing their operations when they use IT systems that can adapt to new environments and follow instructions based on company strategy. A further advantage of AI-based autonomic  computing is server consolidation, which decreases the need for expensive and labour-intensive human oversight of massive server farms   \cite{A33}. Employing AI for autonomic  computing could make handling systems simpler. Consequently, computing systems will see a substantial improvement. Server load distribution is an additional case that illustrates how applications function. It involves dividing tasks among many servers. Additionally, monitoring energy generation closely is crucial for implementing policies that are both environmentally friendly and economical.

Autonomic computing has undergone the following transformations due to AI:
\begin{itemize}

\item \textbf{Economical: }There are benefits to using computing systems rather than physical data centers \cite{A34}. Cloud computing makes it easy for businesses to purchase AI technologies, despite the hefty upfront expenditures. AI systems may potentially do data analysis independently of humans.

\item \textbf{Autonomous: } AI-driven cloud computing has the potential to make businesses smarter, more effective, and more insight-driven   \cite{A35}. Automation of mundane and routine tasks and the ability to undertake data analysis independently of human intervention are two ways in which AI-driven could increase output.

\item \textbf{Data Management:} By combining AI with Google Cloud Stream statistical analysis, it is possible to accomplish real-time customisation, identifying anomalies, and administration scenario forecasting   \cite{A36}.

\item \textbf{Effective Decisions:} The increasing development of cloud applications underscores the importance of intelligence-based data security   \cite{A37}. Security solutions for networks that use AI make it feasible to trace and analyse network traffic. Systems driven by AI can immediately initiate an alarm response upon detection of an anomaly. This method protects important data.

\end{itemize}

\subsection{AI-driven Autonomic Computing: A Conceptual Model}
\begin{figure*}[t]
	\centering
	\includegraphics[scale=0.6]{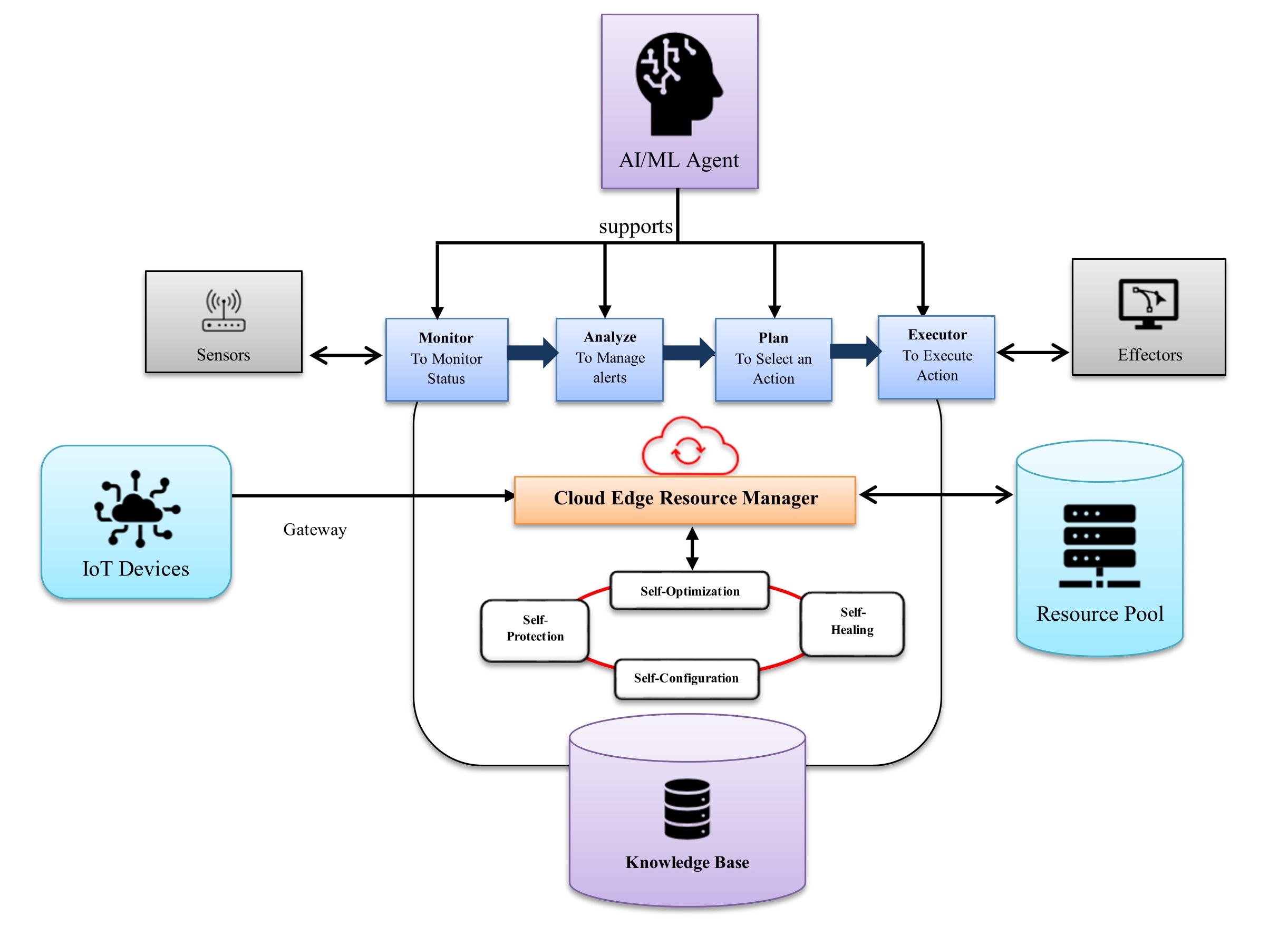}
	\caption{ A Conceptual Model for AI-driven Autonomic Computing \cite{A26}}
	\label{fig8}
\end{figure*}

Figure \ref{fig8} illustrates the conceptual software design, highlighting the connection between AI/ML and the autonomic  computing paradigm \cite{A26}. To meet demands for a wide range of IoT apps, this model incorporates cutting-edge technology to provide efficient computing solutions.

\textbf{1. IoT Applications:} By communicating with IT infrastructures, gateways allow users and IoT gadgets to abstract away interactions with edge-based actuators, sensing devices, and effectors \cite{A26}. To effectively provide AI and other autonomous services, the system can connect to a variety of devices and applications, including medical care, urban planning, agricultural surveillance, and meteorological monitoring.

\textbf{2. Cloud Resource Manager:} Flexible and resilient resource and task allocation is essential for distributed systems, such as edge platforms of IoT. The resource management module tracks the QoS of each distributed task, the needed resources, limitations (such as location), and the available and allocated resources \cite{A26}. In order to produce a workload schedule, the resource manager allocates tasks based on available resources from resource pool and uses AI to estimate their completion time. Using AI and ML approaches, the resource scheduler estimates the system's resources and efficiently allocates them to complete activities within the given time and cost. In addition, this module integrates provider-supplied data on available and planned resources, along with resource descriptions.

\textbf{3. Autonomic Model:} This next-generation architecture makes use of IBM's autonomic computing paradigm with AI/ML, which focuses on capabilities like self-optimization, self-protection, self-configuring, and self-healing. 
\begin{itemize}

\item \textbf{Self-* Properties:} The goal of \textit{self-healing} is to rebound from flaws using the latest AI models while maintaining the whole system functioning without disruption by implementing all necessary alterations. By monitoring unusual behavior and acting accordingly, \textit{self-protection} aims to keep the system safe from malicious intent and ensure its smooth operation. The fundamental objective of \textit{self-configuration} is to install older or missing components automatically using AI, eliminating the need for human intervention \cite{A26}. The \textit{self-optimizing} part uses dynamic scheduling algorithms to assign tasks and workloads to the most suitable resources. AI-driven adaptive scheduling leverages the autonomous component's feedback to consistently enhance the system's efficiency. Data-intensive systems can benefit from AI/ML-based adaptable scheduling due to its adaptability and versatility. In addition, it is possible to automatically assess the effect of various QoS attributes on system efficacy using the latest AI models  \cite{A49}. Additionally, AI-driven autonomic  computing can optimize the distribution of server loads and energy usage, resulting in cost savings and reduced environmental impacts  \cite{A50}. 
\item \textbf{MAPE-k Loop:} \textit{Sensors} record the current state nodes' effectiveness in terms of QoS measures. In order to continuously check for performance differences, \textit{monitors} compare the actual results with those anticipated by AI using data collected from the resource manager node. The monitoring module sends data to the \textit{Analyze and Plan} component, which processes it and determines how to respond based on potential recommendations from AI models. The \textit{executor}'s major goal is self-optimization, which allows it to improve QoS and complete tasks within a certain time frame. The \textit{effector} serves as a conduit for communicating novel rules, guidelines, and warnings to various computing nodes. Along with other QoS needs, the \textit{knowledge base} stores the primary rules for scheduling resources \cite{A26}. AI-powered autonomous agents will assume the role of a system administrator, automating the execution process. Furthermore, AI-driven methods can improve the efficiency of the persistence (knowledge) component of the MAPE-K loop. This is particularly true when it comes to reliably resolving state synchronization in a widely dispersed context.
\end{itemize}

AI-powered autonomic computing has the capability to significantly improve the overall performance of computing systems.

\section{Promising Future Directions}
The following are possible future directions that can be addressed by future readers:
\subsection{Security and Privacy}
Research on IoT, edge computing, and cloud computing has shifted dramatically in recent years toward autonomic computing, all with the goal of better serving consumers \cite{A41}. However, this enormous paradigm shift raises numerous challenges in ensuring the privacy and security of data stored on these devices. For autonomic edge computing to work, privacy and security measures must be strong and flexible. This is because autonomic edge computing has some special features, like less delay, geographical distribution, end-device connectivity, high computational power, fluctuations, and more. Due to the heterogeneity of devices and applications, it is also difficult to develop software platforms that are interoperable  \cite{A42}. The increasing deployment of cloud-based smartphone applications makes intelligence-based autonomous data protection crucial. Security solutions for networks that use AI make it possible to track and analyze network traffic. A \textit{self-protection} module within autonomic computing can use AI to immediately initiate an alarm response upon detection of an anomaly and protect vital data. Applications' security is heavily dependent on their edge infrastructure to withstand threats including replay, denial of service attacks, information degradation, and identification revelation. In autonomic  computing environments, consumers lack control over their personal information, necessitating the establishment of authorization, encryption, and data aggregation. Data on consumers and vendor partners are just one piece of the massive amount of data needed by AI systems  \cite{A43}. Knowing who possesses the data is significantly more valuable than having confidential data that cannot be associated with a specific individual. Companies often face challenges with data security and regulatory compliance when using sensitive content. Privacy and data law security must be considered when implementing AI in autonomic computing \cite{A44}. Given the resource constraints of edge devices of the IoT, AI/ML-powered blockchain computing is the optimal solution for security. These devices simply cannot handle the heavy-duty software and gateways designed for desktop computers. Autonomic computing, which uses AI and ML, might further improve creative software designs that help with the updating and routine maintenance of IoT devices.

\subsection{Edge AI}

Recent advancements in AI effectiveness, the proliferation of IoT devices, and the advent of edge computing have unlocked the potential of edge AI. Because of this, edge AI has potential applications in fields like agricultural fertilization, vehicle assistance, and radiology diagnosis that were previously unthinkable  \cite{A45}. Employing edge AI may improve nearly every industry's current processes. The fact is that cutting-edge applications are propelling the next wave of AI-driven autonomic computing, which will enhance people's lives everywhere—at home, in the office, at school, and even while driving. As a term, "AI at the edge" describes the integration of AI with tangible hardware. "Edge AI" allows the execution of AI computations near users on the network's periphery rather than centralizing all an organization's data in a single location, such as a private data center or a cloud provider's data center  \cite{A46}. Automation is a goal for businesses of all sizes because of the many benefits it provides in terms of efficiency, effectiveness, and security. Software that recognizes patterns and has the ability to consistently perform the same actions could prove beneficial in this situation. The world is unpredictable, and human activities encompass endless conditions, making it challenging to completely communicate them in a system of norms and algorithms. Computerization and devices of any kind may now collaborate with the "intelligence" of human cognition, thanks to advancements in edge AI  \cite{A47}. It is possible for AI-powered intelligent IoT apps to adapt to new environments and efficiently perform the same or comparable duties. Substantial development in crucial areas has enabled the realistic implementation of AI models at the edge. Advancements in neural networks and other branches of AI have also made universal ML possible. Businesses are increasingly realizing that they can efficiently train AI models and then implement them at the edge  \cite{A26}. AI at the peripheral requires distributed computing resources. Graphics Processing Units (GPUs), which have recently achieved massive parallelism, power modern AI-driven autonomic systems. The proliferation of IoT-connected gadgets has led to an unprecedented spike in data volume in modern times. The advancement of data collection devices such as sensors, smart cameras, and robotics has enabled the use of AI-driven autonomic computing at the edge in almost every aspect of business  \cite{A36}. Enhanced reliability, speed, and privacy brought to the battlefield by 5G/6G are equally beneficial to IoT applications.

\subsection{Energy Efficiency and Carbon Neutrality} 

In recent years, there has been a meteoric growth in the amount of data collected and processed. Because of this trend, edge systems have been operating at or near their computational and, consequently, energy consumption limits. Decentralised computing systems like Fog and Edge have emerged as a result of this change \cite{A27}. By offloading some of the processing to dispersed edge devices and networks, autonomic cloud computing is able to significantly reduce latency and costs. Apart from the power source, an irregular energy supply presents significant challenges for a wide range of mission-critical and sensing applications. The exponential growth in the variety of IoT devices and the data they generate has challenged the capacity of networking to process, calculate, and transmit data. Similarly, a shortage of computational power, data storage, and electricity currently hinders the development of smaller IoT devices. Therefore, improving the network's fog and edge nodes' efficiency is crucial. There has been a recent surge in the importance of data centers' sustainability and reducing their carbon footprints \cite{A49}. Therefore, it is crucial to maintain QoS throughout the process. Despite the setbacks, there have been numerous successes. Researchers employ AI based software, hardware, and transitional methods to address the problem of energy management. Computational frameworks support the development of AI based approaches and procedures that maximize software efficiency. Computing at the edge of a network is one example of this. When it came to hardware, the goal was to minimize power consumption while maximizing performance, especially for this application. Several studies have focused on how to make sensor networks more energy efficient. In the preliminary stage, the researchers employed proactive resource management using AI, scheduling for fog/edge-node hibernation times, and other energy-saving tactics. Regarding the effectiveness and longevity of fog, edge, and cloud platforms, there are numerous unresolved issues and potential avenues for future enhancements \cite{A50}. As a result of global warming, the paradigms for autonomic systems that will replace current ones will most likely prioritize electricity and carbon footprints. Systems that assume a constant and steady power supply, a connection to renewable energy sources and alternative methods of reducing energy consumption face more fundamental issues; these problems go beyond the current focus on reducing power consumption per unit of computation. Reduced power usage is necessary to run the CPU and transmitter on energy from the sun or another source. Therefore, reducing the granularity of the fog/edge network could lead to more dispersed, dominant, and robust designs. Studies involving blockchain algorithms and different forms of flexible AI-based autonomic computing could improve energy scheduling in a number of domains, including those dealing with energy limitations.

\section{Conclusions and Summary}
The article delves into the relationship between SLA violations and QoS in cloud-based autonomic computing systems, as well as their respective roles. To learn how QoS needs affect self-management, one must be familiar with the development of autonomic  cloud computing in order to ascertain the scheduling efficiency of the provided resources. Ultimately, it identified unanswered questions about autonomic cloud computing and other open research topics, which ML and AI systems could potentially address by anticipating resource needs. Importantly, self-adaptive systems driven by AI offer to be environmentally friendly and economically address evolving needs in an unpredictable environment, rather than merely allocating increasing amounts of resources. Consequently, some industry heavyweights are studying and implementing autonomic computing with cutting-edge AI and ML solutions.

\end{document}